# Pressure-induced orbital reordering in Na$_2$CuF$_4$

Craig I. Hiley,[1] Catriona A. Crawford,[1] Craig L. Bull,[2,3] Nicholas P. Funnell,[2] Urmimala Dey,[4,5] Nicholas C. Bristowe,[4] Richard I. Walton[1,*] and Mark S. Senn[1,†]

[1]*Department of Chemistry, University of Warwick, Gibbet Hill Road, Coventry, CV4 7AL, United Kingdom.*
[2]*STFC ISIS Facility, Rutherford Appleton Laboratory, Oxfordshire, OX11 0QX, United Kingdom.*
[3]*School of Chemistry, University of Edinburgh, David Brewster Road, Edinburgh EH9 3FJ, Scotland, United Kingdom.*
[4]*Centre for Materials Physics, Durham University, South Road, Durham, DH1 3LE, United Kingdom.*
[5]*Luxembourg Institute of Science and Technology (LIST), Avenue des Hauts-Fourneaux 5, L4362, Esch-sur-Alzette, Luxembourg.*
*Contact author: r.i.walton@warwick.ac.uk
†Contact author: m.senn@warwick.ac.uk

The high-pressure behaviour of Na$_2$CuF$_4$ is explored by powder neutron diffraction and density functional theory (DFT) calculations. A first-order phase transition is observed to take place between 2.4–2.9 GPa, involving a reorientation of the Jahn-Teller (JT) long axes of the [CuF$_6$] octahedra (and therefore the $d_{z^2}$ Cu orbitals), in agreement with our DFT calculations which suggest a transition at ~2.8 GPa. The transition can be described as being between a state of ferro-orbital order and one of A-type antiferro-orbital order, reflecting a shift in the associated electronic instability from being in the zone-center to zone boundary of the first Brillouin zone of the parent structure, with pressure. This change results in a decoupling of magnitude of the associated Jahn-Teller distortion of the Cu-F bond lengths from the lattice strain. This scenario is supported by our observations that the compressibility of the pre-transition phase is highly anisotropic, whilst in the post-transition phase it becomes almost isotropic, and that we observed no further decrease of the magnitude the JT distortion up to 5 GPa, or melting of the OO in our DFT calculations up to at least 5 GPa.

## I. INTRODUCTION

The degenerate electronic ground state of Cu$^{2+}$ ($d^9$) in an octahedral crystal field makes it Jahn-Teller (JT) active, which is manifested as a distortion of the Cu local environment. Many crystalline phases with infinitely connected structures containing Cu$^{2+}$ display a co-operative JT effect, i.e. an ordering of the Cu $d$ orbitals, leading to a lowering of the structure's symmetry relative to its aristotype. For example, whilst the majority of the fluoride perovskites (general formula $AM$F$_3$) are considered to be cubic at room temperature, KCuF$_3$ adopts a tetragonal [1-4] (or orthorhombic [5]) structure due to an elongation of the $c$-axis, yielding one-dimensional $S$ = ½ antiferromagnetism [6], and similarly cuprospinel (CuFe$_2$O$_4$) has a tetragonal structure that is a distorted variant of the spinel structure [7]. The co-operative JT effect has been suggested as the underlying cause of a variety of unusual electronic phenomena such as superconductivity in cuprates [8] and colossal magnetoresistance in manganites [9].

Although the (doped) copper oxide $n$ = 1 Ruddlesden-Popper (RP) phases are the eponymous high-$T_c$ superconducting materials [10], the phenomenon appears to be absent in RP copper fluorides (general formula $A_2$CuF$_4$). Despite being isostructural and isoelectronic, the magnetic interactions are completely different: La$_2$CuO$_4$ displays antiferromagnetic order [11-13] whilst $A_2$CuF$_4$ ($A$ = K, Rb, Cs) all display two-dimensional ferromagnetism in the $ab$ plane [14, 15]. This is possibly on account of small structural distortion present in La$_2$CuO$_4$ consisting of tilting of the [CuO$_6$] octahedra caused by the large difference in ionic radii between $A^+$ and La$^{3+}$ (e.g. nine-coordinate $r_\text{K}$ = 1.55 Å compared to $r_\text{La}$ = 1.216 Å [16]), clearly leads to significant differences in the electronic structure.

Whilst Na$_2$CuF$_4$ has the RP stoichiometry, and $r_\text{Na}$ = 1.24 Å is very close to that of $r_\text{La}$, it adopts a monoclinic structure ($P2_1/c$) consisting of one-dimensional chains of edge-sharing [CuF$_6$] octahedra [17, 18]. A similar structure is adopted by several other 3$d$ metal fluorides which have been considered analogues of the silicate Mg$_2$SiO$_4$, thought to be found at extremely high-pressures [19] and identified as possible Na-ion battery cathode materials [20]. A recent study on Na$_2$CuF$_4$ observed several phase transitions up to 30 GPa by Raman spectroscopy in a diamond anvil cell [19], with the first occurring at 4 GPa, but direct characterisation of the structure and symmetry of these phases is missing in the literature. Motivated by this, and the prospect of transforming Na$_2$CuF$_4$ to the denser RP structural type, we have conducted variable-pressure neutron powder diffraction (which is particularly sensitive to sodium and fluoride ions) up to 5 GPa revealing a structural phase transition driven by orbital reordering.

## II. EXPERIMENTAL TECHNIQUES

NaF (Acros Organics, 97%) and $CuCl_2 \cdot 2H_2O$ (Honeywell, ≥99.0%) and ethylene glycol (Fisher, ≥99%) were all used as received without further purification. Polycrystalline samples were synthesised using a solvothermal method. 1.705 g $CuCl_2 \cdot 2H_2O$ (0.01 mol) and 1.260 g NaF (0.03 mol) were added to 100 ml ethylene glycol. The resulting solution was added to a 200 ml Teflon-lined steel autoclave and heated to 120 °C for 120 hrs. After cooling, the sample was filtered and rinsed with a small additional quantity of ethylene glycol and subsequently allowed to dry at 70 °C overnight (1.145 g yield).

The sample was initially characterised by laboratory X-ray powder diffraction (XRD) using a Panalytical Empyrean diffractometer with Cu K$\alpha_{1,2}$ (1.5406 Å, 1.5444 Å) source. Refinements against laboratory XRD data showed a small (0.78(14) wt%) impurity of NaF, as well as some unidentified even weaker peaks (Figure S1). No attempt was made to remove the NaF by washing with water, since it was found in preliminary experiments that $Na_2CuF_4$ readily reacts with water to form Cu(OH)F.

Time-of-flight variable-pressure neutron diffraction measurements [21] were carried out using the PEARL diffractometer [22] at the ISIS Neutron and Muon Facility. A small amount of sample (~0.2 g) was encapsulated in a TiZr gasket with a Pb pressure marker and deuterated 4 : 1 methanol and ethanol mixture as pressure transmitting medium. The gasket was sealed between a pair of single-toroidal profiled anvils machined from zirconia-toughened alumina (ZTA) and the assembly was loaded into a V3 Paris-Edinburgh press. An initial load of 4 tonnes was applied, and then increased in 2 tonne increments (~0.2 GPa) up to 20 tonnes (with a resulting pressure of ~2 GPa), followed by 5 tonne increments (~0.5 GPa) up to 60 tonnes (with a resulting pressure of ~ 5GPa). The applied load was then decreased to 5 tonnes and then 0 tonnes. Resulting diffraction patterns showed peaks from the ZTA anvils ($ZrO_2$, $Al_2O_3$) and the Pb pressure marker, which were included as phases in Rietveld refinement models. All structural refinements were carried out against powder diffraction data using Topas version 6 [23] (further details of Rietveld refinements against variable-pressure neutron diffraction data are included in Results and Discussion). The contribution of the secondary phases (NaF and an unidentified phase that was visible by XRD (see above)) to the variable-pressure neutron diffraction patterns collected using PEARL with lower signal : noise from reasonable collection times is minimal: the diffraction peaks from NaF were not observed and only weak unidentified diffraction peaks were visible at $Q$ = 2.75 Å$^{-1}$ and 2.85 Å$^{-1}$ (Figure S2). To simplify the model, these additional peaks were not accounted for in Rietveld refinements against variable-pressure powder neutron patterns and were treated as part of the background. The applied pressure at the sample was determined from the refined Pb lattice volume using the Birch-Murnaghan equation of state [24] (Equation S1) for Pb [25].

The structural models of the high-pressure phase and the aristotypes were generated using functions within the ISOTROPY suite [26] (further details form part of the Results and Discussion). The software PASCal [27] was used to calculate the principal compressibilities and carry out a Birch-Murnaghan equation of state fit against the refined cell volume of $Na_2CuF_4$.

First-principles calculations were performed within the density functional theory (DFT) framework as implemented in the VASP code [28, 29], version 6.3.2. PBEsol variant of the generalised gradient approximation (GGA) was used as the exchange-correlation functional, which is known for its accuracy in describing bulk structural properties [30]. PAW pseudopotentials (PBE, potpaw.64) [31, 32] with the following valence configurations were employed: $2s^22p^63s^1$ (Na), $3p^64s^13d^{10}$ (Cu), and $2s^22p^5$ (F). Convergence tests performed on a 14-atom unit cell of the ambient-pressure monoclinic structure revealed that a plane wave energy cutoff of 900 eV and a $k$-mesh grid of 9 × 3 × 5 were necessary to resolve the total energy within 0.002 meV/f.u. The energy convergence criterion was set to $10^{-9}$ eV and performed full relaxations until the Hellmann–Feynman forces on each atom converged to 1 meV/Å. Correlation effects of the $3d$ electrons of Cu were included using $U$ = 7.5 eV and $J$ = 1.0 eV [19, 33-35] within the DFT+$U$ approach introduced by Dudarev et al. [36].

## III. RESULTS AND DISCUSSION

Rietveld refinement against powder neutron diffraction data at $p$ = 0.193(4) GPa shows good agreement with the monoclinic ($P$ 1 $2_1/c$ 1) structure of $Na_2CuF_4$ determined by single crystal X-ray diffraction [18] (Figure 1a, Table I), named herein as $Na_2CuF_4$-I. Upon application of pressure, the onset of a phase transition is observed at $p$ = 2.44(2) GPa which is not complete until 3.39(3) GPa (Figure 1b). This coexistence of both the initial and new phases is indicative of a first-order phase transition. No further phase transitions were observed up to $p$ = 5.05(7) GPa, and upon depressurisation to ambient pressure the material returns to the $Na_2CuF_4$-I structure (Figure S3).

Whilst the $Na_2CuF_4$-I ($P$ 1 $2_1/c$ 1) model provides an approximate fit to the high-pressure diffraction data; it does not capture all features with some peaks poorly fitted (Figure S4b, Table S1). None of the previously calculated structures proposed by Upadhyay et al. [19] provided a satisfactory fit to our high-pressure diffraction data, nor did the related orthorhombic structure adopted by $Na_2FeF_4$ [20] (see Figure S4, Table S1). An orthorhombic ($Ammm$) [37] structural aristotype ($a$ = 3.2621 Å, $b$ = 9.3607 Å, c = 5.6066 Å, Cu1

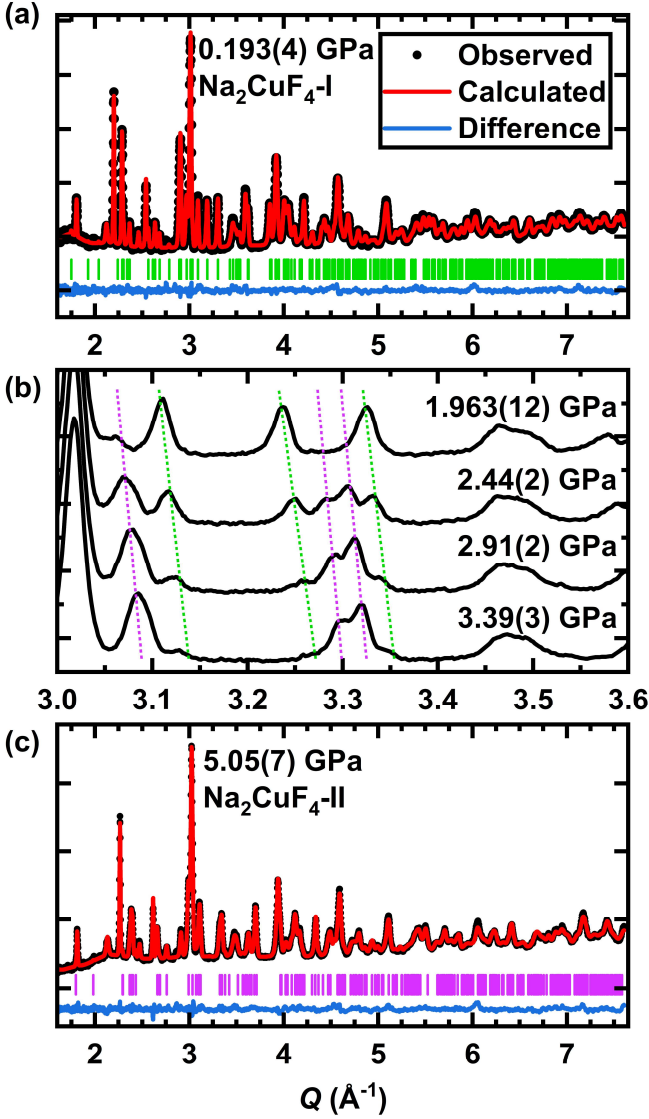

FIG. 1. Variable-pressure neutron powder diffraction data measured at PEARL plotted in scattering vector, $Q$ ($=2\pi/d$). (a) Rietveld refinement of the Na$_2$CuF$_4$-I ($P\,1\,2_1/c\,1$) structure against data measured at $p = 0.193(4)$ GPa. (b) Diffraction patterns measured around the phase transition with an incremental offset for clarity. As a guide for the eye, coloured lines track the approximate positions of diffraction peaks of the Na$_2$CuF$_4$-I (green) and Na$_2$CuF$_4$-II (purple) phases. (c) Rietveld refinement of the Na$_2$CuF$_4$-II ($P\,1\,1\,2_1/b$) structure to data measured at $p = 5.05(7)$ GPa. Tick marks in (a) and (c) are only for the Na$_2$CuF$_4$ phase with tick marks from the sample environment and pressure marker phases omitted for clarity.

(0,0,0), F1 (½,0,¼), F2 (0,¼,0), Na (½,¼,¼)) was identified using the FINDSYM software [38] and permitted subgroups were identified by ISODISTORT [39], both of which are part of the ISOTROPY software suite [26]. The $Ammm$ aristotype consists of one-dimensional chains of edge-sharing [CuF$_6$] octahedra extending along the $a$-axis direction, separated by Na ions (Figure 2b). The space group symmetry confines all Cu–F1 bonds (which form the edge-sharing network) to have the same length, precluding any kind of long-range orbital ordering driven by JT distortions along the chains.

TABLE I. Refined parameters for Na$_2$CuF$_4$ polymorphs from Rietveld refinements to neutron powder diffraction at $p = 0.194(3)$ GPa and 5.05(7) GPa.

|  | Na$_2$CuF$_4$-I | Na$_2$CuF$_4$-II |
|---|---|---|
| Pressure / GPa | 0.194(3) | 5.05(7) |
| $R_{wp}$ / % | 3.03 | 3.68 |
| Space group | $P\,1\,2_1/c\,1$ | $P\,1\,1\,2_1/b$ |
| $a$ / Å | 3.26249(11) | 3.1695(2) |
| $b$ / Å | 9.3613(3) | 9.0986(5) |
| $c$ / Å | 5.6071(3) | 5.4685(4) |
| Unique angle / ° | $\beta$, 92.441(2) | $\gamma$, 90.777(6) |
| $V$ / Å$^3$ | 171.091(12) | 157.686(19) |
| Cu1, 2$a$ | 0, 0, 0 | |
| Na1, 4$e$ | 0.5217(17), 0.1858(7), 0.4136(8) | 0.499(2), 0.1830(9), 0.4193(9) |
| F1, 4$e$ | 0.5692(9), 0.4502(4), 0.2689(7) | 0.5799(14), 0.4522(5), 0.2628(7) |
| F2, 4$e$ | 0.0109(12) 0.1861(5) 0.1268(7) | 0.0131(15), 0.1929(6), 0.1294(7) |
| $B_{iso}$ | 0.79(5) | 0.19(5) |

Additionally, the chains can display no rotation around the chain axis (equivalent to the crystallographic $a$-axis).

Octahedral rotation around the chain axis ($Y_2^+$ distortion mode, where the rotation direction alternates between neighbouring chains) lowers the symmetry to $Pmcb$ (Figure 2c). From $Pmcb$, applying either the $Y_4^+$ or $Y_3^+$ lead to either Na$_2$CuF$_4$-I ($P\,1\,2_1/c\,1$) or to a structure with the space group $P\,1\,1\,2_1/b$ (Na$_2$CuF$_4$-II), respectively (Figure 2a), that provides a very good fit to the high-pressure neutron powder diffraction data at 5.05(7) GPa (Figure 1c). While both distortions lead to a monoclinic cell, each has an additional degree of freedom (that transform at the $\Gamma_3^+$ and $\Gamma_4^+$ irreproducible representations (irreps), respectively), and lead to a different unique axis. The $Y_4^+$ (in Na$_2$CuF$_4$-I) and $\Gamma_4^+$ (in Na$_2$CuF$_4$-II) distortion modes displace the "apical" (i.e. not part of the edge-sharing network) F2 site along the $a$-axis from the aristotypical site. On the other hand, the $\Gamma_3^+$ (as in Na$_2$CuF$_4$-I) and $Y_3^+$ (as in Na$_2$CuF$_4$-II) modes correspond to $x$-direction displacement of the F1 site which allows for two Cu–F1 bond distances, allowing the JT elongated ($l$) Cu–F bond to lie within the edge-sharing network in both phases. The key structural difference between Na$_2$CuF$_4$-I and Na$_2$CuF$_4$-II lies in the orientation of the $l$ Cu–F bonds. In Na$_2$CuF$_4$-I, the $\Gamma_3^+$ distortion leads to $l$ Cu–F bonds being parallel within the chains and pseudo-parallel (i.e. parallel if the $Y_2^+$ chain rotation is disregarded) to those of the nearest neighbouring chain (Figure 2d). In Na$_2$CuF$_4$-II, the $Y_3^+$ distortion still yields parallel interchain $l$ Cu–F bonds, but the intrachain $l$ Cu–F bonds are now pseudo-perpendicular (Figure 2e). Since one of these distortions may be viewed as transforming as a zone-center and the other as a zone-boundary irrep of $Ammm$, the transition can be considered as going from ferro-orbital (Na$_2$CuF$_4$-I) to A-type antiferro-orbital (Na$_2$CuF$_4$-II, with ferro-orbital ordering in one-dimension, and antiferro-orbital ordering between nearest neighbours in two dimensions [40]) ordering. The relative energies of each structure at ambient pressure were calculated by DFT, and it was found that Na$_2$CuF$_4$-II is only ~2 meV/f.u. less stable than Na$_2$CuF$_4$-I (Figure 2b-e), compared to the intermediate symmetry phase ($Pmcb$), which is over 50 meV/f.u. higher. With the exception of the $Ammm$ aristotype, the structures were all predicted to have $A$-type antiferromagnetic ground states (Figure S5, Table S2).

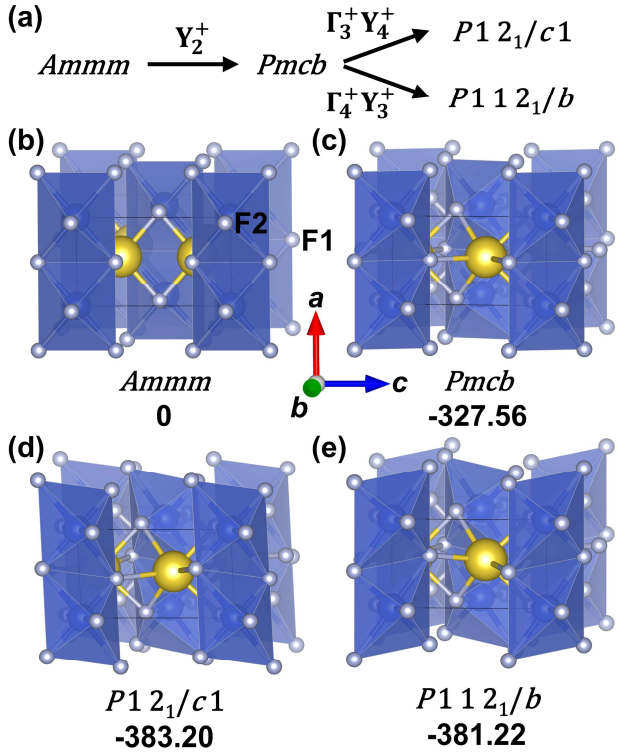

**FIG. 2.** (a) Space group relationship showing the derivation of the Na$_2$CuF$_4$-I ($P\,1\,2_1/c\,1$) and Na$_2$CuF$_4$-II ($P\,1\,1\,2_1/b$) structures from the aristotype ($Ammm$) via an orthorhombic intermediate ($Pmcb$). (b-e) the crystal structures of the aristotypes and observed phases with DFT calculated energies of the relaxed structures (in meV/f.u.), relative to the aristotype at ambient pressure. Sodium atoms are shown in yellow with blue [CuF$_6$] octahedra. The same cell directions are used throughout (N.B. the monoclinic structures have unique axis ≈ 90°).

The structures of Na$_2$CuF$_4$-I and Na$_2$CuF$_4$-II allow a good fit to the data at all pressure points studied (up to $p = 5.05(7)$ GPa, Figure 1c). Rietveld fits against each dataset were carried out (in addition to phases arising from the sample environment and pressure marker) using: only Na$_2$CuF$_4$-I from $p = 0.193(4)$ GPa to $p = 1.963(12)$ GPa; both Na$_2$CuF$_4$-I and Na$_2$CuF$_4$-II at $p = 2.44(2)$ GPa and $p = 2.91(2)$ GPa; and only Na$_2$CuF$_4$-II from $p = 3.93(3)$ GPa to $p = 5.05(7)$ GPa. To minimise the number of refined parameters, a single atomic displacement parameter was refined across all sites. Plots of refined lattice parameters as a function of pressure are available in Figure S6a-c.

In Na$_2$CuF$_4$-I, $\beta$ tends towards 90° as the pressure increases from 92.441(2)° at $p = 0.193(12)$ GPa to 91.65(2)° at $p = 2.91(2)$ GPa. However, in Na$_2$CuF$_4$-II the oblique angle $\gamma$ is largely invariant with pressure at ~90.8° between $p = 2.44(2)$ GPa and $p = 5.05(7)$ GPa (Figure 3a). In both phases the change in cell lengths remain continuous and the oblique angle's proximity to 90° means its variability has a small influence on cell volume, $V$, leading to no significant discontinuity in $V$. To demonstrate this, a single 3$^{rd}$ order Birch-Murnaghan equation of state [24] (Equation S1) was fitted to $V$ of both Na$_2$CuF$_4$-I and Na$_2$CuF$_4$-II as a function of pressure using PASCal (Figure 3b, Table II) [27], with a bulk modulus, $B_0$, of 46.1(4) GPa. In the absence of a significant decrease in $V$, it is not immediately obvious what is driving this subtle pressure-induced phase transition, since the local bonding environments of each atom is essentially unchanged (Figure S6d-f).

**TABLE II.** Birch-Murnaghan equation of state (Equation S1) parameters for Na$_2$CuF$_4$ (both polymorphs) for a 3$^{rd}$ order fit from PASCal. $B_0$ is bulk modulus, $V_0$ is the unit cell volume at zero pressure and $B'$ is the pressure derivative of the modulus.

| $B_0$ (GPa) | $V_0$ (Å$^3$) | $B'$ | $R^2$ |
|---|---|---|---|
| 46.1(4) | 171.82(2) | 5.4(3) | 0.9998 |

The $\Gamma_1^+$ distortion mode (that transforms as a breathing mode of the [CuF$_6$] octahedron) which is has the same variation with pressure as the cell volume (Figure S7a). The distortion mode $Y_2^+$ (transforming as the rotation of the octahedra around the chain axis, Figure 2c) remains almost constant as a function of pressure, even between the two phases (Figure S7b). This results in chains of octahedra that are rotated by ~45° relative to those in their neighbouring chain (determined by measurement of the angle between the apical Cu–F bond directions in neighbouring chains) in both phases at all pressures measured. The $Y_4^+$ and $\Gamma_4^+$ modes (relating to $x$-displacement of the apical F2 site relative to the aristotype) in Na$_2$CuF$_4$-I and Na$_2$CuF$_4$-II, respectively, are small but show systematic decrease with pressure in Na$_2$CuF$_4$-I with little variation as a function of pressure in Na$_2$CuF$_4$-II (Figure 3c). The $\Gamma_3^+$ and $Y_3^+$ modes correspond with F1 $x$-displacement in Na$_2$CuF$_4$-I and Na$_2$CuF$_4$-II, respectively, also display little dependence on pressure but have a much greater magnitude than the modes affecting the F2 site (Figure 3c). The deviation of the $\beta$ and $\gamma$ angles from 90° can be represented as symmetry breaking strain modes $\Gamma_3^+$ / $\Gamma_4^+$ (respectively) which are proportional to the amplitudes of the atomic displacements transforming as the same irreps (Figure S8).

The distortion of the [CuF$_6$] octahedron, $\rho_0$, is quantified using Van Vleck $Q_2$ and $Q_3$ distortion modes (Equations S2-5) [41]. As a function of $p$, $\rho_0$ in the Na$_2$CuF$_4$-I decreases linearly, i.e. the octahedron is becoming less distorted as pressure increases, concordant with other Cu phases (Figure 3d) [42, 43]. Around the phase transition $\rho_0$ in Na$_2$CuF$_4$-II at 3.39(3) GPa is unchanged (within one standard deviation) compared to $\rho_0$ in Na$_2$CuF$_4$-I at 1.963(12) GPa, and thereafter remains approximately constant upon the application of increasing pressure, indicating Na$_2$CuF$_4$-II is better able to retain the remaining JT distortion upon the application of pressure (up to at least 5.05(7) GPa). A polar plot ($\varphi = \tan^{-1}(Q_2/Q_3)$) shows that the character of the JT distortion remains unchanged with pressure and corresponds to an almost pure $Q_3$ 2-long bond: 4-short bond distortion, as is typically observed for Cu$^{2+}$ compounds [44] (Figure S9). Thus, it appears both the character and magnitude of the JT distortion is robust to the application of hydrostatic pressure in Na$_2$CuF$_4$-II. This is in marked contrast to what is observed in canonical JT distorted fluoride KCuF$_3$ (and to a lesser extent, Na$_2$CuF$_4$-I) which shows a gradual suppression of $\rho_0$ upon compression over the same pressure range [44], see Figure S9. It would therefore appear the phase transition from to Na$_2$CuF$_4$-II acts to relieve the melting of the orbital ordering in Na$_2$CuF$_4$-I.

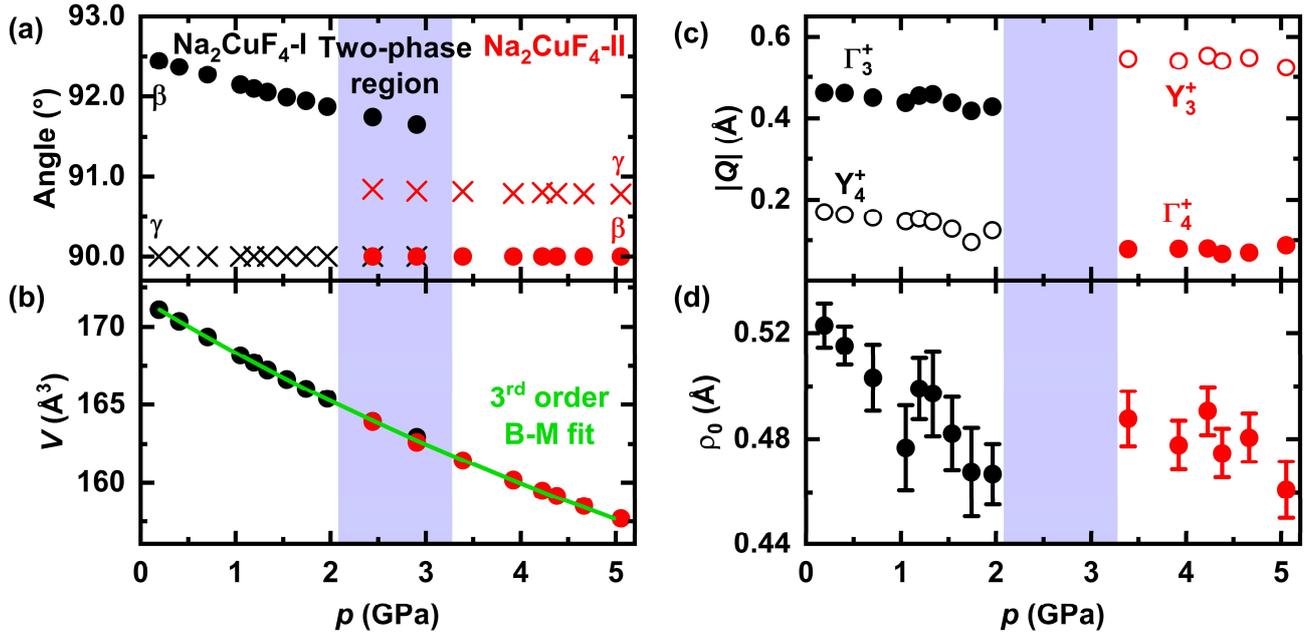

FIG. 3. Structural parameters from Rietveld fits against variable-pressure neutron powder diffraction data. Where not shown, error bars are smaller than data points. (a) Unique angles in Na$_2$CuF$_4$-I ($P\,1\,2_1/c\,1$, black) and Na$_2$CuF$_4$-II ($P\,1\,1\,2_1/b$, red) phases. γ and β in Na$_2$CuF$_4$-I and Na$_2$CuF$_4$-II, respectively are also shown. (b) Cell volume, $V$, with 3$^{rd}$ order Birch-Murnaghan equation of state fit to $V$ in both phases shown in green. (c) Distortion mode amplitudes, $|Q|$, for modes related to distortion of [CuF$_6$] octahedra. (d) Octahedral distortion, $\rho_0$. In (c) and (d), the values obtained in the two-phase region were highly correlated due to peak overlap and have been omitted.

To investigate this point further, the compressibility of Na$_2$CuF$_4$ was calculated using the software PASCal [27]. It was necessary to calculate the compressibilities of Na$_2$CuF$_4$-I and Na$_2$CuF$_4$-II separately, with the former studied up to 1.963(12) GPa and the latter studied from 2.44(2) GPa. PASCal first calculated a set of orthogonalised principal axes for each phase with principal directions denoted $X_i$ and $X'_i$ for Na$_2$CuF$_4$-I and Na$_2$CuF$_4$-II respectively (Figure 4a-c). $X_2$ aligns precisely along the crystallographic $b$-axis, whilst $X_1$ and $X_3$ lie within the $ac$ plane at approximately 45° to the $a$- and $c$-axes. $X'_3$ lies precisely on the $c$-axis, whilst $X'_1$ and $X'_2$ align approximately along $a$- and $b$-axes, respectively. The relative change in principal axes' lengths, $\ell_i$, for both phases show good agreement with empirical fits, with $R^2 > 0.999$ for all axes (Figure 4d). From these empirical fits, the principal compressibilities, $K_i$ and $K'_i$ (for Na$_2$CuF$_4$-I and Na$_2$CuF$_4$-II, respectively) and can be calculated as a function of pressure (Figure 4e). In Na$_2$CuF$_4$-I, $X_1$, which approximately aligns with the direction of the $l$ Cu–F bond (Figure 4b), is the most compressible direction, suggesting that the $l$ bond undergoes the most compression, corresponding to the decrease in the octahedral distortion, $\rho_0$ of the Cu in the ambient pressure phase as a function of pressure (Figure 3d). $X_3$, which is approximately aligned with the short Cu–F1 bond, is the least compressible. In Na$_2$CuF$_4$-II, the compressibilities are much more isotropic, explaining the relative invariance of $\rho_0$ to pressure in this phase. Notably, $K'_1$ demonstrates a small increase in compressibility as a function of pressure (i.e. the structure becomes softer along the chain direction as the pressure increases). In Na$_2$CuF$_4$-II the $l$ Cu–F bonds (Figure 4c) are (pseudo)-perpendicular, meaning they are no longer all aligned along a single principal axis, but alternate in the $X'_1X'_3$ plane. The phase transition thus acts to decouple the orbital ordering from the hydrostatic pressure-induced compression of the unit cell by shifting the electronic instability from the Brillouin zone center to a zone boundary.

The energy of Na$_2$CuF$_4$-II relative to Na$_2$CuF$_4$-I was computed as a function of pressure using DFT in steps of 0.5 GPa from 0 to 5 GPa. At each pressure, the structure was allowed to relax, yielding minimised structures with unit cell volumes within 2.5% of those experimentally determined (Figure S10). The energy difference, $\Delta E$, at each pressure is shown in Figure 5. Na$_2$CuF$_4$-II is initially higher in energy than Na$_2$CuF$_4$-I, but becomes more stable as the pressure increases, with a calculated critical pressure, $p_c$, of ~2.8 GPa, compared to the observed $p_c$ of 2.4 GPa – 2.9 GPa.

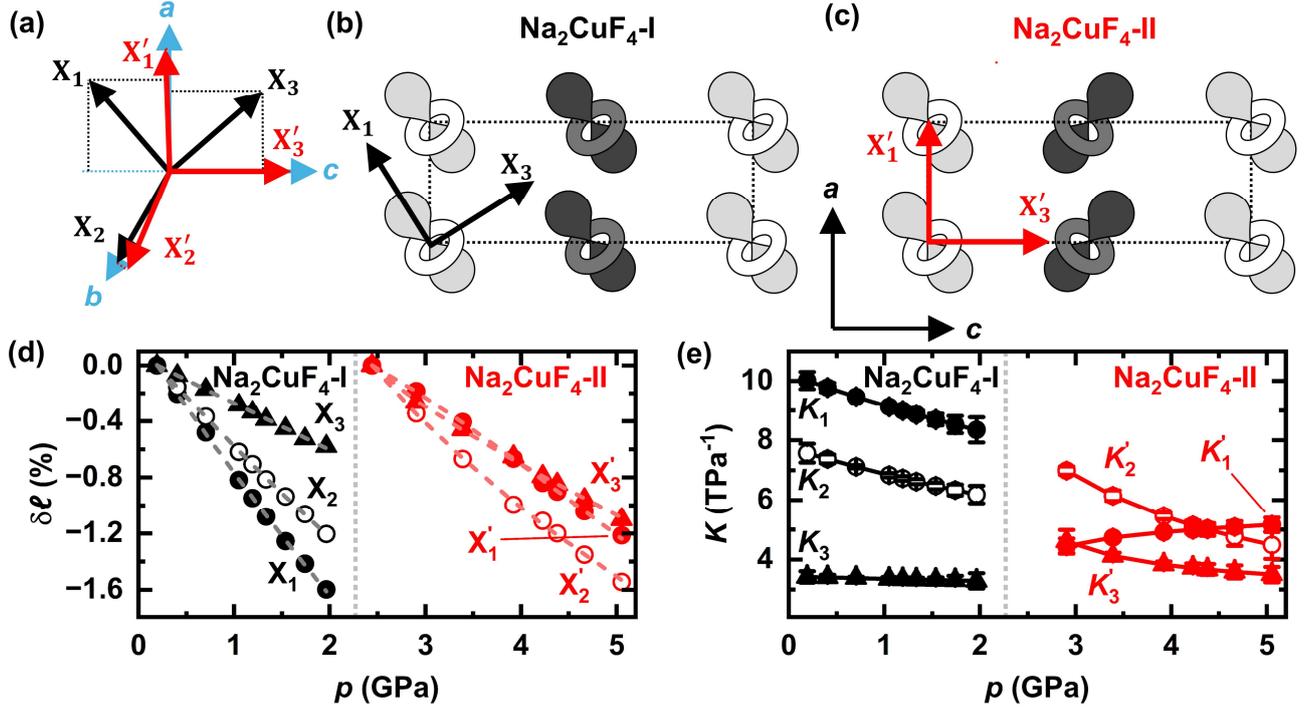

**FIG. 4.** (a) The orthogonalized principal axes of Na$_2$CuF$_4$-I ($X_i$, black) and Na$_2$CuF$_4$-II ($X'_i$, red) overlaid onto the approximate crystallographic axes (blue). Note, only $X_2$ and $X'_3$ are precisely aligned to a crystallographic axis. Exact principal axis vectors are given in Tables S3 and S4. (b,c) Schematic of Na$_2$CuF$_4$-I and Na$_2$CuF$_4$-II (respectively) Cu $d_{z^2}$ orbital ordering viewed (approximately) along the $b$-axis, with principal axis directions overlaid. Light-shaded orbitals are from Cu atoms on the cell origin whilst darker-shaded are from Cu atoms with a y fractional coordinate of 0.5. (d) The % change in principal axes' lengths $\delta\ell$ as a function of pressure (relative to their initial lengths at 0.193(4) GPa and 2.44(2) GPa for $X_i$ and $X'_i$, respectively) with empirical fits (dashed lines, parameters given in Tables S3 and S4). (e) The derived principal compressibilities, $K$, (data points are joined by solid lines for clarity).

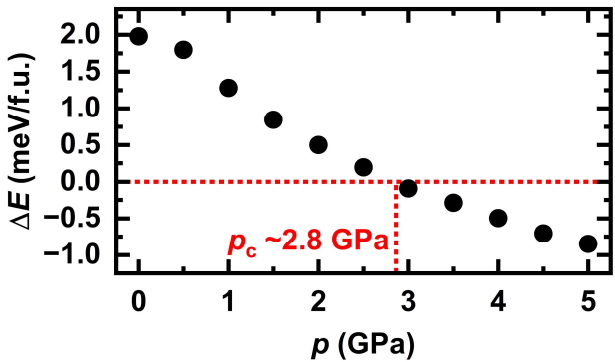

**FIG. 5.** Energy of the high-pressure Na$_2$CuF$_4$-II phase relative to the ambient-pressure Na$_2$CuF$_4$-I phase ($\Delta E$) as a function of pressure, $p$, with the calculated $p_c$ indicated.

There are other Cu-containing crystal structures that display orbital re-ordering under pressure. For example, in the hybrid perovskite [(CH$_3$)$_2$NH$_2$]Cu(HCOO)$_3$, a re-ordering occurs at 5.2 GPa, caused by a change in the degree of octahedral tilting [45]. High-pressure structural studies of copper fluoride RP phases have found that in (K,Rb)$_2$CuF$_4$ an orbital re-ordering occurs at ~9 GPa such that the JT elongated Cu–F bonds realign from along the $c$-axis to along the $a$-axis disrupting the ambient pressure ferromagnetic structure [46, 47], whilst at 2.2 GPa Cs$_2$CuF$_4$ begins to display antiferromagnetic next-nearest-neighbour intralayer ordering whilst retaining its interlayer ferromagnetism [48], suggesting a change in the orbital ordering. In Na$_2$CuF$_4$, the magnetic interactions are unlikely to be drastically different in the two polymorphs, since the one-dimensional edge-sharing [CuF$_6$] chains that constitute the strongest magnetic super-exchange pathways are essentially unchanged. Furthermore, the quasi-1-dimensionality of both structures suppresses their magnetic order [49], making elucidation of these interactions extremely difficult. The high-pressure behaviour of Na$_2$CuF$_4$ reported here is relatively unusual because the magnitude of the orbital ordering instability remains approximately constant, whilst the point at which it occurs is switching from the Brillouin zone center to a zone boundary, leading to a transition from ferro- to A-type antiferro-orbital ordering.

## IV. CONCLUSIONS

The crystal structure of $Na_2CuF_4$ as a function of pressure has been studied by powder neutron diffraction up to 5 GPa. At 2.4 – 2.8 GPa a first-order phase transition occurs from $P\ 1\ 2_1/c\ 1$ to $P\ 1\ 1\ 2_1/b$, consisting of a reorientation of half the Cu–F JT long ($l$) bonds to give interchain pseudo-perpendicular $l$ bonds, compared to pseudo-parallel $l$ bonds seen in $Na_2CuF_4$-I, i.e. a transition from ferro-orbital to A-type antiferro-orbital ordering. We find this transition is driven by the ability of $Na_2CuF_4$-II to accommodate JT $l$ Cu–F bonds at high pressure, as evidenced by a constant octahedral distortion factor ($\rho_0$). The transition can be viewed as a switch in the point at which an instability arises in the Brillouin zone, from the zone center to a zone boundary. DFT calculations show that $Na_2CuF_4$-II becomes more stable at $p_c \sim 2.8$ GPa, in agreement with the experimental observations.

## ACKNOWLEDGEMENTS

R.I.W. and M.S.S. acknowledge the Leverhulme Trust for a research project grant (Grant No. RPG-2022-22). U.D. and N.C.B. acknowledge the Leverhulme Trust for a research project grant (Grant No. RPG-2020-206). M.S.S. acknowledges the Royal Society for a fellowship (UF160265 URF\R\231012). We thank the U.K. Science and Technology Facilities Council for access to the PEARL instrument at ISIS Neutron and Muon Facility (experiment RB2410566). This work has made use of the Hamilton HPC Service of Durham University. The initial sample characterisation was performed via the Warwick X-ray Research Technology Platform.

# Supplemental Material for "Pressure-Induced Orbital Reordering in Na$_2$CuF$_4$"

**Craig I. Hiley,[1] Catriona A. Crawford,[1] Craig L. Bull,[2,3] Nicholas P. Funnell,[2] Urmimala Dey,[4,5] Nicholas C. Bristowe,[4] Richard I. Walton[1] and Mark S. Senn.[1]**

[1]Department of Chemistry, University of Warwick, Gibbet Hill Road, Coventry, CV4 7AL, United Kingdom.

[2]STFC ISIS Facility, Rutherford Appleton Laboratory, Oxfordshire, OX11 0QX, United Kingdom.

[3]School of Chemistry, University of Edinburgh, David Brewster Road, Edinburgh EH9 3FJ, Scotland, United Kingdom.

[4]Centre for Materials Physics, Durham University, South Road, Durham, DH1 3LE, United Kingdom.

[5]Luxembourg Institute of Science and Technology (LIST), Avenue des Hauts-Fourneaux 5, L4362, Esch-sur-Alzette, Luxembourg.

**Supplemental Figures**

**Supplemental Tables**

**Supplemental Figures**

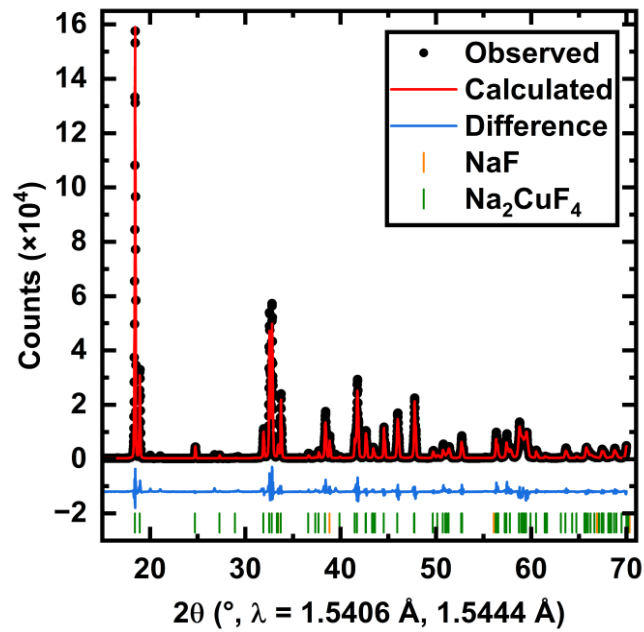

*Figure S1. Rietveld refinement against laboratory XRD for $Na_2CuF_4$-I with 0.78(14) wt% NaF.*

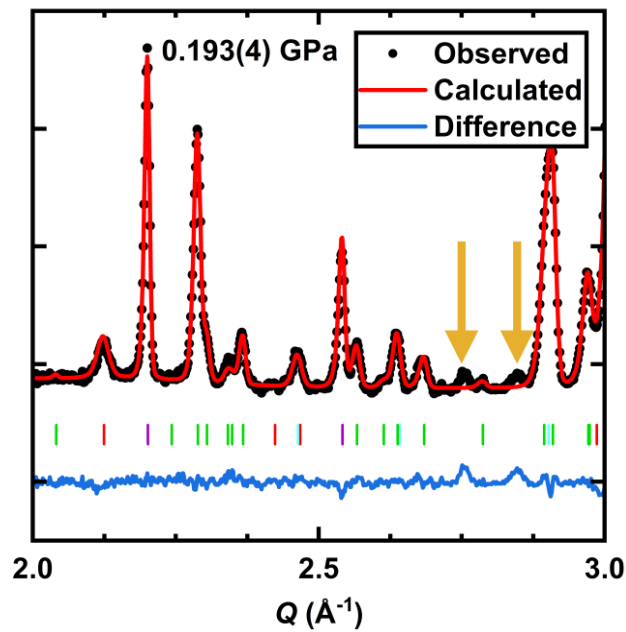

*Figure S2. Neutron diffraction pattern collected at p = 0.193(4) GPa from $Na_2CuF_4$-I inside variable-pressure assembly with unfitted peaks at Q = 2.75 Å$^{-1}$ and 2.85 Å$^{-1}$ highlighted with orange arrows. Tick marks denote peak positions of $Na_2CuF_4$-I (green), $Al_2O_3$ (cyan), $ZrO_2$ (red) and Pb (purple).*

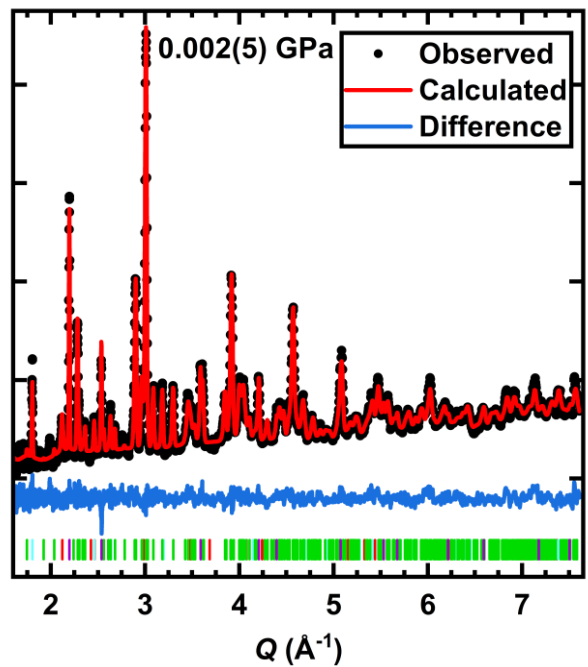

*Figure S3. Neutron diffraction patterns upon depressurisation to 0.002(5) GPa. Tick label colours are consistent with Figure S2.*

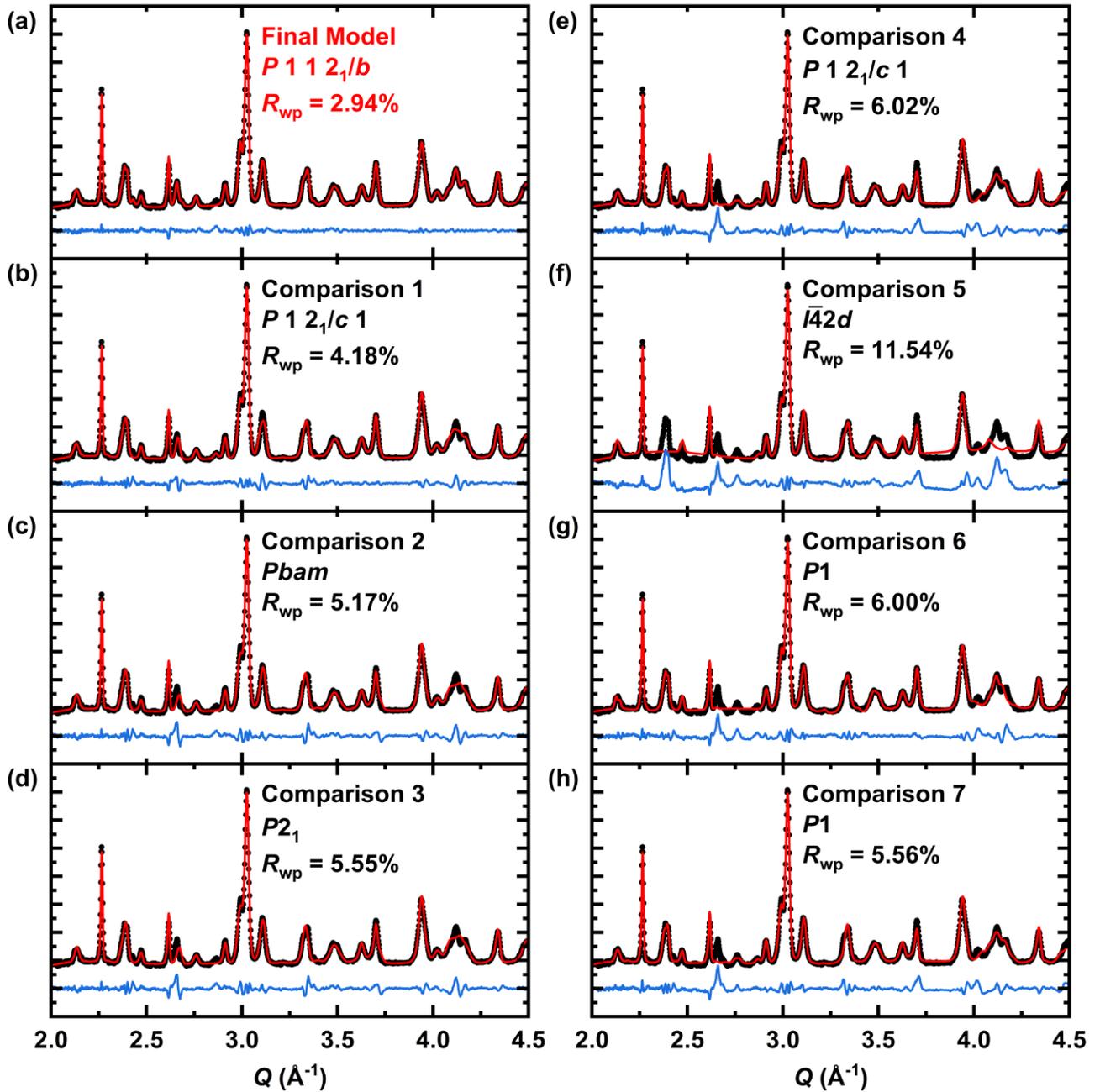

*Figure S4. Fits against neutron diffraction data measured at* p *= 5.05(7) GPa using the Rietveld method for sample environment and pressure marker, and Pawley fits for the following structural models (a) $Na_2CuF_4$-II, (b) $Na_2CuF_4$-I, (c) orthorhombic cell for $Na_2FeF_4$ [1], (d-h) high-pressure phases calculated by DFT by Upadhyay* et al. *[2]. Further details of fits in Table S1. The fits were carried out over the* Q*-range ~1.5 Å$^{-1}$ to ~7.6 Å, but only the region between 2 Å$^{-1}$ and 4.5 Å$^{-1}$ are plotted for clarity.*

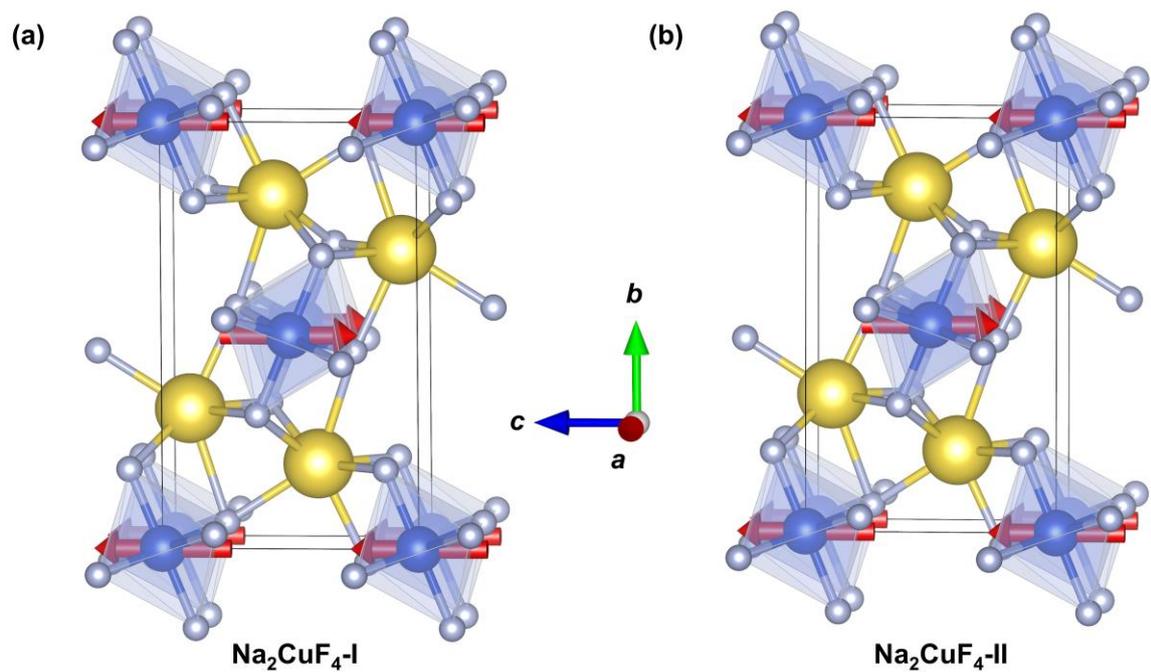

*Figure S5. A-type antiferromagnetic ground states of (a) Na$_2$CuF$_4$-I and (b) Na$_2$CuF$_4$-II calculated by DFT (see Table S2).*

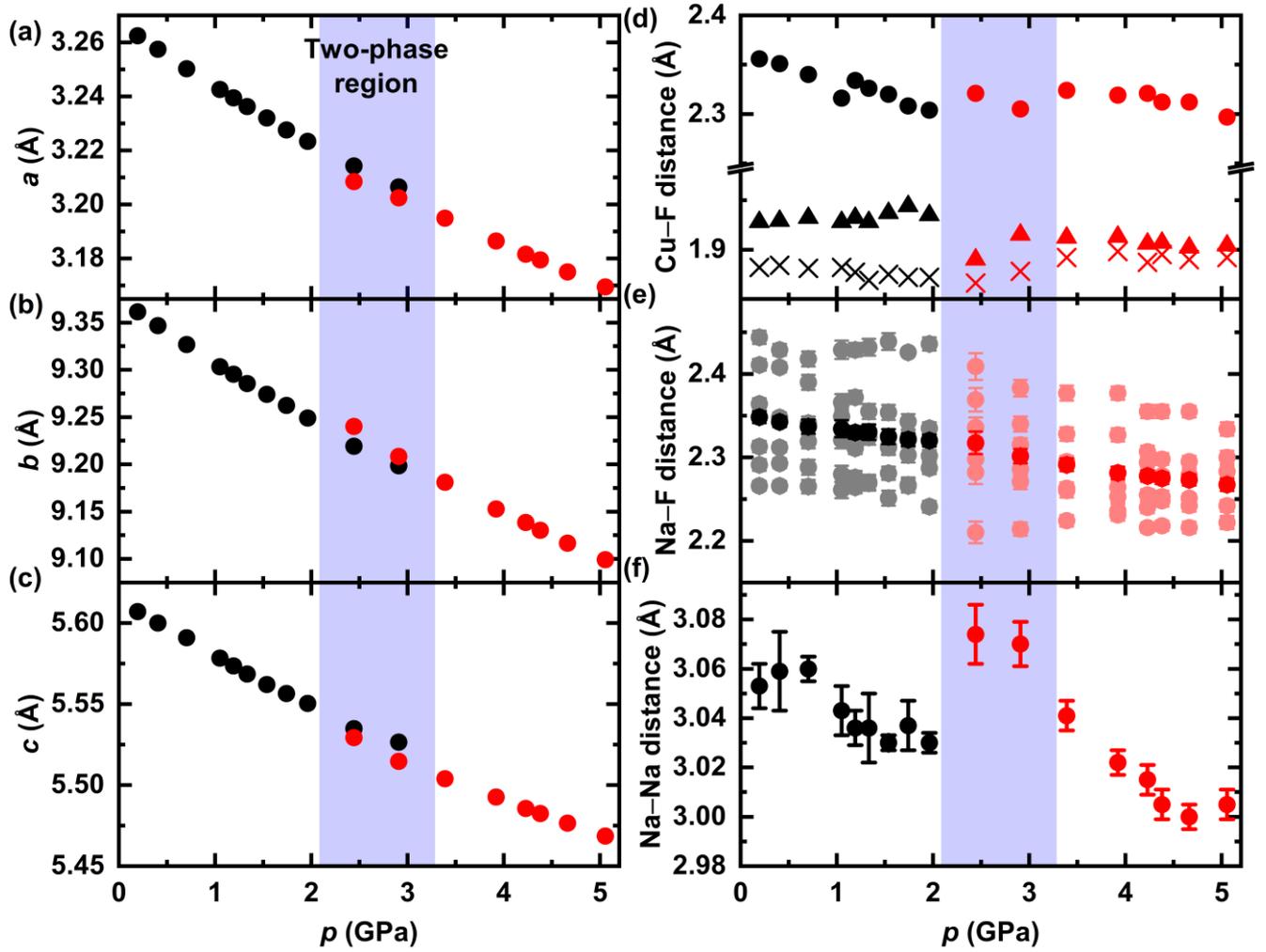

*Figure S6. (a-c) Refined lattice parameters of $Na_2CuF_4$-I (black) and $Na_2CuF_4$-II (red) as a function of pressure. (d,e) Refined metal–F distances (mean Na–F distances shown in e) and (f) closest Na–Na distance.*

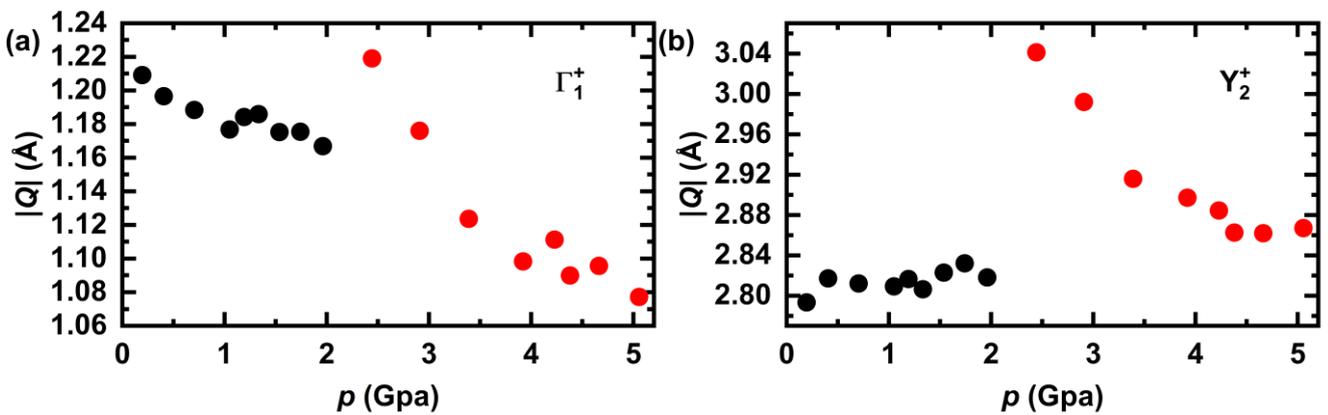

*Figure S7. Refined distortion mode amplitudes, |Q| of (a) $\Gamma_1^+$ and (b) $Y_2^+$ as a function of pressure for $Na_2CuF_4$-I (black) and $Na_2CuF_4$-II (red).*

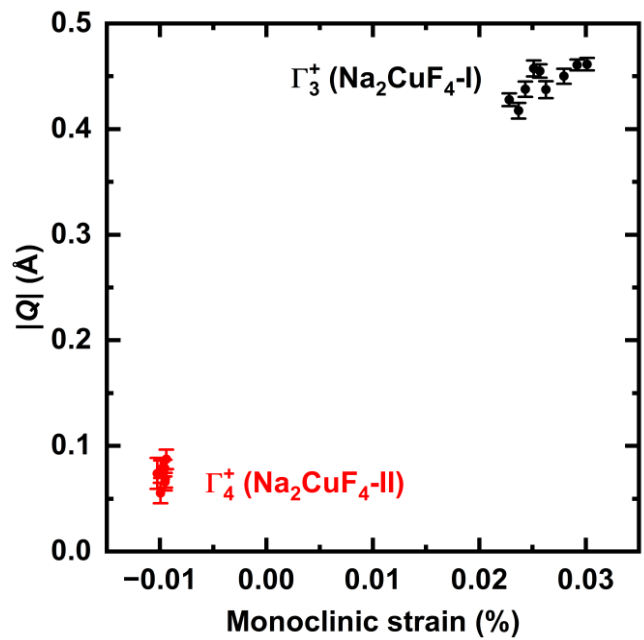

*Figure S8. Distortion mode vs monoclinic strain for $\Gamma_{3,4}^+$.*

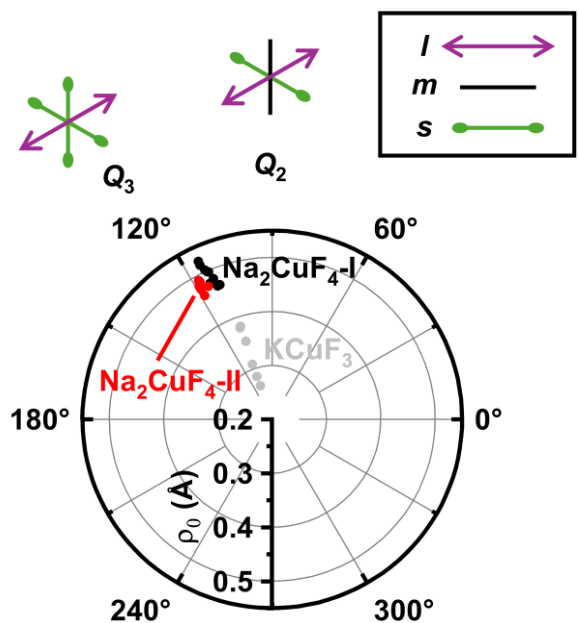

*Figure S9. Polar plot of with a radial magnitude $\rho_0$ (Equation S4) and angle $\varphi$ (Equation S5) for $Na_2CuF_4$-I, $Na_2CuF_4$-II between 0.194(3) GPa and 5.05(7) GPa, and $KCuF_3$ between 0 and 5 GPa (calculated from bond lengths tabulated by Zhou et al. [3]), where* l *= long,* m *= medium and* s *= short length bonds.*

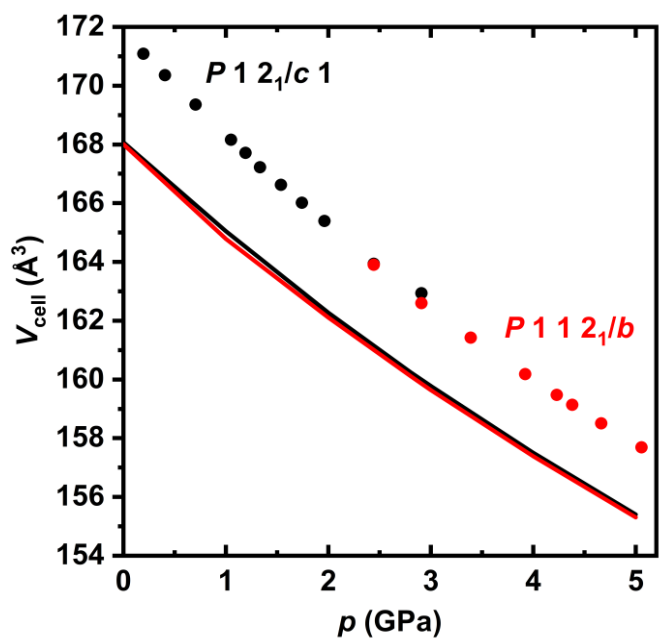

*Figure S10. Comparison of observed cell volumes (points) to those calculated by DFT (lines) as a function of pressure. The $Na_2CuF_4$-I and $Na_2CuF_4$-II phases are represented as black and red, respectively.*

## Supplemental Tables

*Table S1. Lattice parameters for $Na_2CuF_4$ Pawley models against neutron diffraction data measured at 5.05(7) GPa (see Figure S4). Lattice parameters with no uncertainty reported were not refined, as dictated by symmetry or model (for 6 and 7).*

|   | Space group | a / Å | b / Å | c / Å | α / ° | β / ° | γ / ° | $R_{wp}$ / % | Ref. |
|---|---|---|---|---|---|---|---|---|---|
| - | $P112_1/b$ | 3.1697(4) | 9.0978(5) | 5.4698(8) | 90 | 90 | 90.789(8) | 2.94 | This work |
| 1 | $P12_1/c1$ | 3.1726(11) | 9.0946(10) | 5.4768(15) | 90 | 90.42(2) | 90 | 4.18 | [4] |
| 2 | $Pbam$ | 5.4582(12) | 9.0969(11) | 3.1746(7) | 90 | 90 | 90 | 5.17 | [1] |
| 3 | $P2_1$ | 3.1024(10) | 9.0611(18) | 5.4200(13) | 90 | 102.886(13) | 90 | 5.55 | |
| 4 | $P12_1/c1$ | 3.1047(9) | 9.0558(19) | 5.4187(12) | 90 | 102.921(12) | 90 | 6.02 | |
| 5 | $I\bar{4}2d$ | 5.9404(11) | 5.9404(11) | 6.4420(12) | 90 | 90 | 90 | 11.54 | [2] |
| 6 | $P1$ | 5.851(3) | 5.955(8) | 6.295(3) | 90 | 89.12(8) | 89.55(5) | 6.00 | |
| 7 | $P1$ | 5.822(8) | 6.058(14) | 6.286(10) | 89.57(15) | 89.19(16) | 90 | 5.56 | |

*Table S2. DFT-calculated relative energies of the considered magnetic configurations for each structural model.*

| Structure | Relative Energies (meV/f.u.) | | |
|---|---|---|---|
| | Ferromagnetic | A-Type Antiferromagnetic | C/G-Type Antiferromagnetic |
| $Ammm$ | -0.011795 | 0.00 | -1.8449725 |
| $Pmcb$ | 0.1491225 | 0.00 | 0.2838125 |
| $P\,1\,2_1/c\,1$ | 0.1130375 | 0.00 | 0.23509 |
| $P\,1\,1\,2_1/b$ | 0.0677875 | 0.00 | 0.2319175 |

*Table S3. Parameters outputted by PASCal [5] fit after input of lattice parameters of Na$_2$CuF$_4$-I (0.194(3) GPa ≤ p ≤ 1.963(12) GPa). Principal axis with median compressibility (K), the principal axis' projection onto the unit cell parameters and the empirical parameters used to fit change in principal axis relative length (see Figure 4d), according to: $\ell(p) = \ell_0 + \lambda(p - p_c)^v$. The median volume compressibility is also presented.*

| Principal Axis | K (TPa$^{-1}$) | Direction | | | Empirical Parameters | | | |
|---|---|---|---|---|---|---|---|---|
| | | a | b | c | $\ell_0$ | $\lambda$ | $p_c$ | $v$ |
| X$_1$ | 9.00(10) | 0.9005 | 0 | -0.4349 | 5.8011 | -5.6151 | -8.5254 | 0.0151 |
| X$_2$ | 6.72(8) | 0 | 1 | 0 | 4.0006 | -3.881 | -7.6137 | 0.0148 |
| X$_3$ | 3.34(6) | 0.8229 | 0 | 0.5681 | 0.1281 | -0.0249 | -20.1983 | 0.5436 |
| Volume | 19.2(2) | | | | | | | |

*Table S4. Parameters outputted by PASCal [5] fit after input of lattice parameters of Na$_2$CuF$_4$-II (2.44(2) GPa ≤ p ≤ 5.05(7) GPa). Principal axis with median compressibility (K), the principal axis' projection onto the unit cell parameters and the empirical parameters used to fit change in principal axis relative length (see Figure 5d), according to: $\ell(p) = \ell_0 + \lambda(p - p_c)^v$. The median volume compressibility is also presented.*

| Principal Axis | K (TPa$^{-1}$) | Direction | | | Empirical Parameters | | | |
|---|---|---|---|---|---|---|---|---|
| | | a | b | c | $\ell_0$ | $\lambda$ | $p_c$ | $v$ |
| X$_1'$ | 5.46(11) | -0.4058 | 0.9139 | 0 | 0.0443 | -0.0364 | 0.6381 | 0.332 |
| X$_2'$ | 3.83(11) | 0 | 0 | 1 | 0 | -0.0049 | 2.4431 | 0.8397 |
| X$_3'$ | 4.92(8) | 0.9983 | 0.0584 | 0 | 0 | -0.0044 | 2.4431 | 1.0869 |
| Volume | 14.7(3) | | | | | | | |

**Supplementary Information**

The 3$^{rd}$ order Birch-Murnaghan equation of state [6] used in Figure 3b, with corresponding fit parameters in Table 2:

$$p = \frac{3B_0}{2}(\eta^7 - \eta^5)\left[1 + \frac{3}{4}(B' - 4)(\eta^2 - 1)\right], \quad \text{(Equation S1)}$$

where $\eta = (V_0/V)^{1/3}$.

The $Q_2$ and $Q_3$ Van Vleck modes used to express the distortion of the CuF$_6$ octahedron were calculated according to the equations expressed by Kanimori [7]:

$$Q_2 = l - s, \quad \text{(Equation S2)}$$

$$Q_3 = \frac{(2m - l - s)}{\sqrt{3}}, \quad \text{(Equation S3)}$$

where *l*, *m* and *s* are the long, medium and short bond lengths, respectively. A polar plot can be created with a magnitude of $\rho_0$ versus an angle $\varphi$:

$$\rho_0 = \sqrt{(Q_2^2 + Q_3^2)}, \quad \text{(Equation S4)}$$

$$\varphi = tan^{-1}\left(\frac{Q_2}{Q_3}\right), \quad \text{(Equation S5)}$$